\def\a{\alpha}\def\e{\epsilon}
\def\f{\phi}
\def\m{\mu}\def\n{\nu}\def
\p{\pi}\def\q{\psi}\def\r{\rho}\def\t{\tau}

\def\O{\Omega}\def\S{\Sigma}

\def\de{\partial}\def\na{\nabla}
\def\inf{\infty}\def\mo{{-1}}\def\ha{{1\over 2}}
\def\qu{{1\over 4}}

\def\gmn{g_{\m\n}}\def\ghmn{\hat g_{\m\n}}\def\mn{{\mu\nu}}  
\def\ds{ds^2=}

\def\af{asymptotically flat }\def\st{spacetime }
\def\bh{black hole }
                                   
\def\bg{background }\def\gs{ground state }\def\bhs{black holes }

\def\ct{conformal transformation }
\def\gi{gravitational instanton }

\def\ads{anti-de Sitter }
\def\RN{Reissner-Nordstr\"om }

\def\BR{Bertotti-Robinson }\def\MP{Majumdar-Papapetrou } 
\def\GR{general relativity }

\def\section#1{\bigskip\noindent{\bf#1}\smallskip}
\def\nota{\footnote{$^\dagger$}}

\def\CMP#1{Commun.\ Math.\ Phys.\ {\bf#1}} 
 
\def\PR#1{Phys.\ Rev.\ {\bf#1}} 
\def\NP#1{Nucl.\ Phys.\ {\bf#1}} 
\def\JMP#1{J.\ Math.\ Phys.\ {\bf#1}}

\def\ref#1{\medskip\everypar={\hangindent 2\parindent}#1}
\def\beginref{\begingroup
\bigskip
\centerline{\bf References}
\nobreak\noindent}
\def\endref{\par\endgroup}

\magnification=1200\baselineskip18pt
\def\ef{e^{-2\f}}\def\km{{(1-k)/2}}\def\kp{{(1+k)/2}}\def\kt{{(3+k)/2}}   
\def\up{\left(1-{r_+\over r}\right)}\def\um{\left(1-{r_-\over r}\right)}
\def\un{\left(1-{r_-\over r_+}\right)}\def\uq{\left(1+{\a\over\r}\right)} 
\def\rp{r_+}\def\rq{r_-}\def\bfx{{\bf x}}\def\bfxi{{\bf x}_i} 
\def\xj{|\bfx-\bfx_j|}\def\ix{|\bfx-\bfx_i|}\def\ij{{\hat\imath\hat\jmath}}

{\nopagenumbers
\line{May 1996\hfil INFNCA-TH9611}
\vskip40pt
\centerline{\bf Multi-black holes and instantons in effective string theory}
\vskip40pt
\centerline{\bf S. Demelio}
\vskip10pt
\centerline {Dipartimento di Scienze Fisiche, Universit\`a di Cagliari}
\centerline{via Ospedale 72, 09124 Cagliari, Italy}
\vskip20pt
\centerline{\bf S. Mignemi}
\vskip10pt
\centerline {Dipartimento di Matematica, Universit\`a di Cagliari}
\centerline{viale Merello 92, 09123 Cagliari, Italy}

\vskip40pt
{\noindent The effective action for string theory which takes into account
non-minimal coupling of moduli admits multi-black hole solutions. The
euclidean continuation of these solutions can be interpreted as an instanton 
mediating the splitting and recombination of the throat of extremal 
magnetically charged black holes.}
\vfil\eject}

\section{1. Introduction}
String theories are presently considered one of the most promising candidates
for a theory of quantum gravity. It is therefore natural that their
implications on the theory of gravity and especially on black holes had
been widely investigated. Since at the moment the formalism of string theories
cannot deal directly with these problems, this has usually been done by means
of field theoretical low-energy effective
actions, which correct the action of \GR by the addition of non-minimally
coupled scalar fields, such as the dilaton or moduli fields.
On the other hand, if the gravitational field is described by a quantum
theory, its topology should fluctuate at Planck length scales [1]. A first
approximation to these effects should therefore be obtained in the context
of a euclidean path integral formalism [2]. In this formalism a fundamental
role is played by finite action regular solutions of the classical field
equations, interpolating between different topological sectors. These
configurations give the main contribution to the path integral in the context
of a semiclassical approximation.

It is then worthwhile to consider this approach in the context of effective
string theory. Unfortunately, however, very little is known about
gravitational instanton solutions of effective string actions [3-4].
In general relativity, an interesting case of \gi is given by the euclidean
multi-\bh solutions in the presence of a Maxwell field. This class of
instantons possesses several asymptotic regions which approach \BR universes.
Brill [5] has exploited this property for the calculation of the probability
of the splitting of a \BR universe into two or more. This is especially
relevant because a \BR universe approximates very well the throat of an
extremal \RN \bh near the horizon and the probability should therefore
approximate the splitting or recombination ratio of  extremal
\RN black holes.

In this paper, we discuss the generalization of the multi-\bh solutions,
with lorentzian or euclidean signature, to the case of a one-parameter
effective string action which includes the non-minimal coupling
of both the dilaton and a modulus field and interpolates between \GR and
the standard GHS model [6]. We show that the euclidean action vanishes for
these instantons, except in the GR limit, according to the fact that
in the string case both the initial
and final states have vanishing entropy.

\section{2. The multi-black hole solutions}
In [7] an effective action for the heterotic string which took into account
the non-minimal coupling to gravity of a modulus field  due to one-loop
threshold effects was introduced. In terms of the string metric, the
action read:
$$I={1\over16\p}\int\sqrt{-g}\ d^4x\ \ef\left[R+4(\na\f)^2-{2\over 3}(\na\q)^2-
\left(1+e^{2(\f-q\q/3)}\right)F^2\right]\eqno(1)$$
where $\f$ is the dilaton, $\q$ a modulus, $F^2$ is the Maxwell field strength
and $q$ a coupling constant.
It was then shown that exact solutions of the classical field equations
can be found if $\ef={q^2\over 3}e^{-2q\q/3}$. In this case, the action 
simplifies to [4]:
$$I={1\over16\p}\int\sqrt{-g}\ d^4x\ \ef\left[R-{8k\over 1-k}(\na\f)^2-
{3+k\over 1-k}F^2\right]\eqno(2)$$
where $k={3-2q^2\over3+2q^2}$ and $-1\le k\le1$. In particular, for $k=-1$,
the action reduces to the standard string action in absence of modulus 
coupling [6], while in the singular limit $k=1$, the dilaton decouples and
one can recover the Einstein-Maxwell theory.
 
The field equations stemming from (2) are:
$$\eqalignno{&\na_\m(\ef F^\mn)=0&(3)\cr
&R={8k\over 1-k}\left[-\na^2\f+(\na\f)^2\right]+{3+k\over 1-k}F^2&\cr
&R_\mn=4{1+k\over 1-k}\na_\m\f\na_\n\f-2\na_\m\na_\n\f+[2(\na\f)^2-\na^2\f]
\gmn+2{3+k\over 1-k}\left[F_{\m\r}F^\r_{\ \n}-\qu F^2\gmn\right]&\cr}$$

These equations admit \bh solutions with magnetic monopole configurations 
of the Maxwell field given by [7]:
$$\ds-\up\um^k dt^2+\up^\mo\um^\mo dr^2+r^2d\O^2$$
$$\ef=\um^\km\qquad\qquad F_\ij={Q\over r^2}\e_\ij\eqno(4)$$
where $\hat\imath$, $\hat\jmath$ $=2,3$.
The two parameters $\rp$ and $\rq$ are related to the magnetic charge $Q$ and
the ADM mass $M$ of the solution by the relations:
$$M=\ha\rp+{3-k\over4}\rq\qquad\qquad Q^2={1-k\over4}\rp\rq\eqno(5)$$
while the temperature $T$ and the entropy $S$ of the \bh are 
$$T={1\over4\p\rp}\un^\kp\qquad\qquad S=\p\rp^2\un^\km\eqno(6)$$
(These formulae correct some errors present in ref. [7]).

Of special interest are the solutions corresponding to extremal \bh with
$\rp=\rq=\a$, i.e. $Q^2=4{1-k\over(5-k)^2}M^2$.
In this limit, the solution (4) can be written as 
$$\ds-\uq^{-(1+k)}dt^2+\uq^2(d\r^2+\r^2d\O^2)$$
$$\ef=\uq^{(k-1)/2}\qquad\qquad F_{ij}={\sqrt{1-k}\over2}{\a\over(\r+\a)^2}
\e_{ij}\eqno(7)$$
where $\r=r-\a$.
These metrics are regular everywhere in the range $0<\r<\inf$.
Besides the horizon at $\r=0$, the $g_{00}$ component of the metric and the
dilaton field become complex for non-integer $k$ and hence the metric cannot
be continued in that region. The surfaces
$\r=0$ are regular horizons placed at infinite spatial distance. However,
the timelike and lightlike geodesics have a finite extent except for $k=-1$,
and hence the metrics are not geodesically complete in general.
For a more detailed discussion on this point, see ref. [8].

Near the horizon at $\r=0$, the metric takes the form of a \BR universe [9], 
namely, it is the direct product of 2-d \ads\st (or flat space if $k=-1$) 
and a 2-sphere of constant radius:
$$\ds-\left({\r\over\a}\right)^{k+1}dt^2+\a^2{d\r^2\over\r^2}+\a^2d\O^2
\eqno(8)$$
with constant magnetic field: $F_\ij={\sqrt{1-k}\over2\a}\e_\ij$.

It is easy to generalize these solutions to the case of many black holes. In
fact, if one inserts into the field equations (3) the ansatz, suggested by (7),
$$\ds-V^{-(1+k)}dt^2+V^2d\bfx\cdot d\bfx$$
$$\ef=V^{(k-1)/2}\qquad\qquad F_{ij}={\sqrt{1-k}\over2}\e_{ijk}\de_kV
\eqno(9)$$
where $\bfx=(x_1,x_2,x_3)$, $i,j=1,2,3$,
one can check that they are satisfied if $\na^2V=0$. In particular, for a
multi-black hole solution, $V$ takes the form:
$$V=c+\sum_{i=1}^N{\a_i\over\ix}\eqno(10)$$

If one requires that the solutions be \af, $c=1$ and for $N=1$ one recovers 
(7).
If $c=0$, instead, one obtains solutions which are asymptotically \BR also
at infinity. The \af solutions describe a distribution of extremal black holes
with masses $M_i={5-k\over4}\a_i$, magnetic charges $Q_i={\sqrt{1-k}\over2}\a_i$,
and dilatonic charges $\S_i={1-k\over4}\a_i$, in equilibrium, and
generalizes the \MP solution of general relativity [10], which is recovered in
the limit $k=1$. The properties of the metric near $\bfx=\bfx_i$ are analogous
to those of the single extremal black holes described above, with $\a=\a_i$.
In particular, the surfaces at $\bfx=\bfx_i$ are event horizons and the
curvature is regular there. 

By duality, one can also obtain electrically charged solutions. In this case,
the metric, the dilaton and the Maxwell field are 
$$\ds-V^{-2}dt^2+V^{1+k}d\bfx\cdot d\bfx$$
$$\ef=V^\km\qquad\qquad F_{0i}={\sqrt{1-k}\over2}\de_i(V^\mo)\eqno(11)$$
where $V$ is given by (10).
These solutions describe a distribution of masses with electric and dilatonic
charge in equilibrium, but display naked singularities at $\bfx=\bfxi$,
except for $k=1$.

The solutions (9) and (11) are also related by a \ct to those discussed by
Shiraishi [11]. In fact, in terms of the "canonical" metric $\ghmn=\ef\gmn$,
the action (2) becomes:
$$I={1\over16\p}\int\sqrt{-\hat g}\ d^4x\left[\hat R-2(\hat\na\hat\f)^2-
e^{-2a\hat\f}\hat F^2\right]\eqno(12)$$
with $a=\sqrt{1-k\over3+k}$, $\f=a\hat\f$, $F=a\hat F$ and line element
$$\ds-V^{-(k+3)/2}dt^2+V^{(k+3)/2}d\bfx\cdot d\bfx\eqno(13)$$
Also in this case, however, the surfaces at $\bfx=\bfxi$ are naked 
singularities, except in the Einstein limit $k=1$.

\section{3. The euclidean continuation}
The analytical continuation of the solutions (9) to imaginary time can be
interpreted as gravitational instantons. In particular, in the case $c=0$,
one obtains the generalization to the string case of the instanton mediating
the splitting and recombination of several \BR universes, introduced by
Brill [5]. As we have seen before, the throat of an extremal magnetically
charged string \bh has the form of a \BR universe, and therefore our
instantons can be considered as an approximation of those mediating the
splitting of extremal black holes. 

Let us proceed to the extension of the Brill instanton to our models. The
euclidean metric is given by
$$\ds V^{-(1+k)}d\t^2+V^2d\bfx\cdot d\bfx\eqno(14.a)$$
with dilaton and Maxwell field
$$\ef=V^{(k-1)/2}\qquad\qquad F_{ij}={\sqrt{1-k}\over2}\e_{ijk}\de_kV
\eqno(14.b)$$
where
$$V=\sum_{i=1}^N{\a_i\over|\bfx-\bfx_i|}$$

The metric is geodesically complete.
When $\bfx$ approaches $\bfx_j$, $V$ approaches ${\a_j\over|\bfx-\bfx_j|}$ and
the metric takes the \BR form in this limit. Similarly, in the limit $\bfx\to
\inf$, $V\to{\a_\inf\over|\bfx|}$, with $\a_\inf=\sum\a_i$, which is also
of the \BR form. Thus the geometry interpolates between the $N+1$ \BR universes
corresponding to its $N+1$ asymptotic regions.

We point out that the euclidean metrics (14.a) are regular without any need to 
impose 
a periodicity in the time coordinate. This is confirmed by the fact that 
extremal black holes (7) have vanishing temperature for any value of $k$. 

In order to calculate the transition probability, one has to evaluate the
euclidean action, which is the dominant term in a semiclassical approximation
of the path integral. This is given by:
$$\eqalign{I_E=&-{1\over16\p}\int_M\sqrt g\ d^4x\ \ef\left[R-{8k\over 1-k}
(\na\f)^2-{3+k\over 1-k}F^2\right]\cr
&-{1\over8\p}\int_{\de M}\sqrt{^3g}\ d^3x\ \ef K}\eqno(15)$$
where $K$ is the trace of the second fundamental form and the boundary term
is needed to ensure the unitarity of the theory.

By using the field equations, the volume integral can be converted into a
surface integral, so that the action can be written:
$$I_E=-{1\over8\p}\int_{\de M}\sqrt{^3g}\ d^3x\ \ef\left[{4k\over 1-k}
n\cdot\na\f+K\right]\eqno(16)$$

For the boundary 3-surfaces, one can choose the cylinder with "mantles"
$|\bfx|=P$, $\t\in(-T,T)$ and $\xj=P_j$, $\t\in(-T,T)$ and basis $|\bfx|\le
P$, $\xj\ge P_j$, $\t=T$ ($-T$), in the limit $P\to\inf$, $P_j\to 0$,
$T\to\inf$ [5]. On the bases the integral vanishes, while on the mantle
surfaces it is given by
$${k-1\over2}P_iT\eqno(17)$$
This vanishes identically for $k=1$, while is undetermined otherwise.
One has therefore to carefully choose a regularization prescription, which can
be fixed by requiring that one obtain zero when the action is evaluated
for the initial, single \BR universe. This can be achieved by taking the
limits $P_i\to 0$, before the limit $T\to\inf$ [5]. 
With this prescription, one obtains a null contribution from the mantle 
surfaces at finite distance.
For $x\to P$, instead, the integral diverges, but the action has to be
renormalized
by subtracting the contribution of the single initial \BR universe, which
exactly cancels the diverging contribution.

The net result is that the action vanishes.
However, there is another contribution to the action coming from the
cylinder two-dimensional edges at $|\bfx|=P$, $\xj=P_j$, $\t=\pm T$,
where the extrinsic curvature has a delta-function behaviour [5].
The contribution to (15) from the $j$th pair of edges is given by [12]:
$$-2{e^{-2\f(P_j)}\over8\p}{\p\over2}A_j=-{\p\over2}P_j^\km\a_j^\kt\eqno(18)$$
where $A_j$ is the area of the edges, given by $4\p\a_j^2$.
One must again subtract the contribution of the initial \BR
universe. The final result is 
$$I_E={\pi\over2}\left(\left(\sum\a_i\right)^2-\sum\a_i^2\right)\eqno(19)$$
for $k=1$, while for $k\ne1$ it vanishes as $P_j^\km$, since it is suppressed
by the dilaton\nota{Some authors [13] have argued that for extremal \bhs the
time coordinate should be periodically identified. In this case, of course, 
the region of integration has no edge and the action is always zero. With 
this prescription, anyway, also the entropy of the $k=1$ extremal black holes
vanishes, in agreement with the thermodynamical argument given below.}.

The action of the instanton is therefore null, except for the GR limit.
Thus, it appears that in the dilaton-modulus gravity, the probabilities
for the splitting of a \BR universe in two or more does not depend on the
parameters $\a_i$.
This is in accordance with the thermodynamics: comparing with 
(6) one sees that the
euclidean action is half the difference of the entropies of the extremal
black holes whose throats are approximated by these \BR universes (and which
vanish except in the GR limit), so that the semiclassical
probability of the transition is equal to the exponential of the difference
of the entropy of the initial and final states, which gives the probability
of a thermodynamical fluctuation. In particular, if $k\ne1$,
the entropy of the extremal black hole vanishes, and there is no potential
barrier to obstruct the process.

\section{4. Final remarks}

In a semiclassical approximation, the probability for the splitting of a
Bertotti-Robin\-son
universe is given at first order by the exponential of the instanton
action $e^{-I_E/\hbar}$ and hence, as first shown in [5], it is suppressed
in the GR limit according to the violation of the second law of thermodynamics,
while, as we have seen, in the other cases the entropy is null both for the
initial and final
states and at this order of approximation there is no potential barrier to
prevent the splitting when the mass of the black holes is of the order of
the Planck scale.

This is interesting in view of the fact that \BR universes approximate the
throat of extremal magnetically charged black holes. As is well known,
extremal \bhs can be considered as the \gs for the Hawking evaporation
process of \bhs of given charge, and from the results obtained one can deduce
that at the last stages of evaporation the extremal string \bhs may split
into smaller ones [14].

It is also interesting to compare the results obtained here with those for
instantons describing pair production of extremal black holes in a \bg
magnetic field [3]. Also in that case, corrections proportional to the
\bh entropy are present in the $k=1$ limit.

\section{Acknowledgements}
{\noindent We wish to thank M. Cadoni for interesting remarks}
\beginref
\ref [1] S.W. Hawking, \NP{B144}, 349 (1978);
\ref [2] S.W. Hawking, in {\it General Relativity, an Einstein centenary
survey}, edited by S.W. Hawking and W. Israel (Cambridge Un. Press 1979);
\ref [3] H.F. Dowker, G. Gauntlett, D.A. Kastor and J. Traschen, \PR{D49}, 958
(1994); H.F. Dowker, G. Gauntlett, S.B. Giddings and G.T. Horowitz, \PR{D50},
2662 (1994); S.F. Ross, \PR{D49}, 6599 (1994);
\ref [4] M. Cadoni and M. Cavagli\`a, \PR{D50}, 6435 (1994);
\ref [5] D.R. Brill, \PR{D46}, 1560 (1992);
\ref [6] D. Garfinkle, G.T. Horowitz and A. Strominger, \PR{D43}, 3140 (1991);
S.B. Giddings and A. Strominger, \PR{D46}, 627 (1992);
\ref [7] M. Cadoni and S. Mignemi, \PR{D48}, 5536 (1993) and \NP{B427}, 669
(1994);
\ref [8] M. Cadoni and S. Mignemi, \PR{D51}, 4319 (1995);
\ref [9] B. Bertotti, \PR{116}, 1331 (1959); I. Robinson, Bull. Akad. Polon.
{\bf 7}, 351 (1955);
\ref [10] S.D. Majumdar, \PR{72}, 930 (1947); A. Papapetrou, Proc. Roy. Irish
Acad. {\bf A51}, 191 (1947); J.B. Hartle and S. Hawking, \CMP{26}, 87 (1972);
\ref [11] K. Shiraishi, \JMP{34}, 1480 (1993);
\ref [12] G. Hayward, \PR{D47}, 3275 (1993);
\ref [13] G.W. Gibbons and R.E. Kallosh, \PR{D51}, 2839 (1995); S.W Hawking,
G.T. Horowitz and S.F. Ross, \PR{D51},4300 (1995) ; C. Teitelboim, \PR{D51},
4315 (1995);
\ref [14] R. Kallosh, A. Linde, T. Ortin, A. Peet and A. van Proeyen,
\PR{D46}, 5278 (1992).
\endref                                                              
\end